\documentclass[twocolumn,showpacs,prb,amsfonts,amsmath,amssymb,floatfix]{revtex4}

\usepackage{hhline}
\usepackage{mathrsfs}
\usepackage{graphicx}
\usepackage{dcolumn}
\usepackage{bm}
\usepackage{color}

\input{epsf}
\begin{document}

\title{
{\it Ab initio} two-dimensional multiband low-energy models of EtMe$_3$Sb[Pd(dmit)$_2$]$_2$ and $\kappa$-(BEDT-TTF)$_2$Cu(NCS)$_2$ with comparisons to single-band models
}

\author{Kazuma Nakamura$^{1,2}$}
\author{Yoshihide Yoshimoto$^{3}$}
\author{Masatoshi Imada$^{1,2}$}
\affiliation{$^1$Department of Applied Physics, University of Tokyo, 7-3-1 Hongo, Bunkyo-ku, Tokyo, 113-8656, Japan} 
\affiliation{$^2$JST TRIP/CREST, 7-3-1 Hongo, Bunkyo-ku, Tokyo, 113-8656, Japan} 
\affiliation{$^3$Department of Applied Physics, Tottori University, Japan} 

\date{August 08, 2012}

\begin{abstract}
We present {\it ab initio} two-dimensional extended Hubbard-type multiband models for EtMe$_3$Sb[Pd(dmit)$_2$]$_2$ and $\kappa$-(BEDT-TTF)$_2$Cu(NCS)$_2$, after a downfolding scheme based on the constrained random phase approximation (cRPA) and maximally-localized Wannier orbitals, together with the dimensional downfolding. In the Pd(dmit)$_2$ salt, the antibonding state of the highest occupied molecular orbital (HOMO) and the bonding/antibonding states of the lowest unoccupied molecular orbital (LUMO) are considered as the orbital degrees of freedom, while, in the $\kappa$-BEDT-TTF salt, the HOMO-antibonding/bonding states are considered. Accordingly,  a three-band model for the Pd(dmit)$_2$ salt and a two-band model for the $\kappa$-(BEDT-TTF) salt are derived. We derive single band models for the HOMO-antibonding state for both of the compounds as well. The HOMO antibonding band of the Pd(dmit)$_2$ salt has a triangular structure of the transfers with a one-dimensional anisotropy in contrast to the nearly equilateral triangular structure predicted in the extended H\"{u}ckel results. The ratio of the larger interchain transfer $t_b$ to the intrachain transfer $t_a$ is around $t_b/t_a \sim 0.82$. Our calculated screened onsite interaction $U$ and the largest offsite interaction $V$ are $\sim$0.7 eV and $\sim$0.23 eV, respectively, for EtMe$_3$Sb[Pd(dmit)$_2$]$_2$ and $\sim$0.8 eV and $\sim$0.2 eV for $\kappa$-(BEDT-TTF)$_2$Cu(NCS)$_2$. These values are large enough compared to transfers $t$ as $\sim$55 meV for the Pd(dmit)$_2$ salt and $\sim$65 meV for the $\kappa$-BEDT-TTF one, and the resulting large correlation strength ($U$$-$$V$)/$t \sim 10$ indicates that the present compounds are classified as the strongly correlated electron systems.  In addition, the validity whether the present multiband model can be reduced to the single-band model for the HOMO-antibonding state, widely accepted in the literature, is discussed. For this purpose, we estimated the order of vertex corrections ignored in the cRPA downfolding to the single band model, which is given by $W'$/$D$, where $W'$ is a full-screened-interaction matrix element between the HOMO-antibonding and other bands away from the fermi level (namely HOMO-bonding or LUMO-bonding/antibonding bands), whereas $D$ is the energy distance between the fermi level and the bands away from the fermi level. In the present materials, $W'$/$D$ estimated as 0.3-0.5 signals a substantial correction and thus the exchange process between the low-energy HOMO-antibonding and other bands away from the fermi level may play a  key role to the low-energy ground state. This supports that the minimal models to describe the low-energy phenomena of the organic compounds are the multiband models and may not be reduced to the single-band model.  
\end{abstract} 
\maketitle 

\section{Introduction}\label{sec:intro}
Searching for possible candidates of a quantum spin liquid state has been one of the central issues in condensed matter physics. Organic materials provide an important research area~\cite{Kanoda,KanodaKato} and a number of organic insulators, especially $\kappa$-(BEDT-TTF)$_2$Cu$_2$(CN)$_3$ (Ref.~\onlinecite{SL-Shimizu}) [where BEDT-TTF is bis(ethylenedithio)-tetrathiafulvalene] and EtMe$_3$Sb[Pd(dmit)$_2$]$_2$ (Ref.~\onlinecite{SL-Itou}) (where dmit is 1,3-dithiole-2-thione-4,5-dithiolate) are strong candidates of the realizations of the spin liquids. Although the nearest neighbor antiferromagnetic couplings are $J$$\sim$220-250 K for both compounds as speculated from the high-temperature magnetic susceptibility data,~\cite{SL-Shimizu,tp-over-t-dmit-1} they do not exhibit long-range magnetic order at least down to the temperature $T$$\sim$5 K. The mechanism of survival of the spin degree of freedom at low temperatures is yet to be clarified. Strong electronic correlation and the resulting large quantum fluctuation are proposed to realize such a ground state,~\cite{Anderson, Moessner, Morita} especially for low-dimensional systems, but it is not so simple to establish and identify the real materials as spin liquids. The low-energy structure of these organic compounds is often described in terms of a single-band model crossing the fermi level. This corresponds to the dimerization limit of the anti-bonding band of the highest occupied molecular orbital (HOMO) of organic molecules forming the dimer. In this limit, the system is described by a triangular-lattice structure and the geometrical frustration in the lattice is considered as a possible origin of the quantum fluctuation. On the other hand, however, an appreciable anisotropy of the transfer structure in the triangular lattice has been reported by several {\it ab initio} density functional calculations.~\cite{NakamuraET,Kandpal,Scriven} In addition, a recent dielectric measurement~\cite{Abdel-Jawad} suggests that the system exhibits a relaxer-type response at low temperatures, $T$$\sim$60 K, thus indicating that the charge degree of freedom might bring about some kinds of disorder in the low-energy electronic state. This is clearly beyond single-band physics.  

Recently, the ground-state properties of EtMe$_3$Sb[Pd(dmit)$_2$]$_2$ are of great interest, for which the extremely low-temperature measurements have been performed. $^{13}$C NMR spectra under 7.65 T do not show significant broadening down to 19.4 mK,~\cite{NMR} while the spin-lattice relaxation rate $T_1^{-1}$ exhibits a sharp drop below 1 K with a temperature dependence $\propto T^2$. It might suggest a continuous phase transition around 1K due to some unknown symmetry breaking accompanied by the existence of gapless excitations in the low-temperature phase. On the other hand, the temperature dependence of heat-capacity\cite{Cp} and thermal-conductivity\cite{thermal} data give a $T$-linear term down to 0.1K, in contradiction with the $T^2$ behavior expected from $T_1^{-1}$. Furthermore, the magnetic torque measurement indicates a Pauli paramagnetic-like uniform magnetic susceptibility down to 30 mK,~\cite{chi} indicating a feature similar to the one-dimensional Mott insulator represented by Heisenberg or Hubbard models at half filling at low temperatures. Such seemingly contradicting experimental data require further careful theoretical analyses on the origin and mechanism of the spin liquid behavior. 

The low-energy electronic structure of EtMe$_3$Sb[Pd(dmit)$_2$]$_2$ is known to be rather unique.  In this material, the level difference between HOMO and lowest unoccupied molecular orbital (LUMO) is small compared to the dimerization gap corresponding to the bonding-antibonding splitting of each orbital~\cite{HOMO-LUMO-gap-1,HOMO-LUMO-gap-2,HOMO-LUMO-gap-3} and the resulting band structure shows unusual level inversion of a pair of the HOMO-LUMO bands.~\cite{HOMO-LUMO-levelcross,Et2Me2Sb-1,Et2Me2Sb-2,Et2Me2Sb-3} Then from the lower energy, the HOMO-bonding (HOMO-b), LUMO-bonding (LUMO-b), HOMO-antibonding (HOMO-ab), and LUMO-antibonding (LUMO-ab) are stacked separated by a finite gap. The HOMO-ab band is half filled and is sandwiched by the LUMO-ab and LUMO-b bands residing above and below 0.5 eV of the fermi level, respectively [see Fig.\ref{Fig1}(a)]. Such a unique band structure may easily provide potential relevance of multiband physics, for instance, charge fluctuations within a dimer corresponding to the polarization between the crossed HOMO-LUMO level.  In addition, a valence-bond-solid phase exists in similar compounds,~\cite{VBStoSC-1,VBStoSC-2,VBStoSC-3} where superconductivity appears in the vicinity. On the other hand, the geometrical frustration in the half-filled HOMO-ab band,~\cite{Kanoda,KanodaKato,tp-over-t-dmit-1,tp-over-t-dmit-2} as well as interactions at a dimer site and/or between dimer sites, can also be relevant to the survival of the spin-liquid phase. To clarify roles of various factors in the low-energy physics, {\it ab initio} derivation of the low-energy model from first principles is highly desired. 

For {\em ab initio} derivations for the low-energy model, a scheme based on the constrained random-phase approximation (cRPA) (Refs.\onlinecite{Aryasetiawan} and \onlinecite{Solovyev}) for constructed maximally localized Wannier orbitals (MLWO) (Ref.\onlinecite{MaxLoc}) has been applied to a wide range of materials.~\cite{ImadaMiyake} Furthermore, by utilizing a quasi-low-dimensional character of materials as organic conductors, the methodology for obtaining {\it ab initio} effective models in reduced dimensions has been developed~\cite{Nakamura2D} and applied to a two-dimensional (2D) effective-model derivation for $\kappa$-(BEDT-TTF)$_2$Cu(NCS)$_2$.~\cite{Shinaoka} In Ref.~\onlinecite{Shinaoka}, a single-band model for the HOMO-ab band crossing the fermi level was derived and analyzed by a multi-variable variational Monte-Carlo method to demonstrate a quantitative reliability of the derived model parameters through the study for the metal-insulator transition. It was found that, while this material is experimentally located in the metallic region with the superconductivity at low temperatures close to the border of the metal-insulator transition,~\cite{NCS-1,NCS-2} the derived {\it ab initio} model and its Monte-Carlo solution predicted an antiferromagnetic insulator. With the 20\% reduction of the interaction parameters, the {\it ab initio} model just gave a metallic solution. As a possible origin of the discrepancy between the theory and experiment, the multiband nature missing in the model construction was discussed, especially in terms of dynamical effects between the target band and the closest high-energy band (in this material, the HOMO-b band). In fact, in the study on the onsite Hubbard model for the $\kappa$-BEDT-TTF system,~\cite{Kuroki} with the fluctuation exchange approximation, the symmetry of the superconducting gap function seems to be sensitive depending on whether the HOMO-b band is included in the degree of the freedom of the effective model or not. 

In the present paper, we derive 2D multiband models of EtMe$_3$Sb[Pd(dmit)$_2$]$_2$ and $\kappa$-(BEDT-TTF)$_2$Cu(NCS)$_2$ from first principles. The single-band models are also derived and compared with the multiband models to understand the essence of the multiband nature. We found that, for EtMe$_3$Sb[Pd(dmit)$_2$]$_2$, onsite intraorbital $U$ and interorbital $U'$ interactions are both $\sim$0.7 eV and the Hund's rule coupling $J$ is 0.1-0.2 eV. An offsite interaction $V$ is $\sim$0.25 eV and a transfer $t$ is 40-50 meV. As a result, our estimated correlation strength ($U$$-$$V$)/$t$ gives a substantially large value as $\sim$10. The other compound $\kappa$-(BEDT-TTF)$_2$Cu(NCS)$_2$ also exhibits large correlation strength as $\sim$10. We also estimate the reliability of the downfolding whether the multiband model can be reduced to the single-band model with the {\it ab initio} cRPA framework. The measure of the reliability is given by an order estimate of the vertex correction to the effective interaction. We found that, for both EtMe$_3$Sb[Pd(dmit)$_2$]$_2$ and $\kappa$-(BEDT-TTF)$_2$Cu(NCS)$_2$, the vertex correction is not small and may well affect the low-energy physics, thus suggesting that the multiband analysis beyond the single-band one is needed for describing the low-energy physics properly.  

The present paper is organized as follows: In Sec. II, we define effective models to be derived and describe a scheme for the derivation. Computational details and results for EtMe$_3$Sb[Pd(dmit)$_2$]$_2$ and $\kappa$-(BEDT-TTF)$_2$Cu(NCS)$_2$ are given in Sec. III. We discuss characteristic aspects of the derived effective models including correlation strength and the estimate of the reliability of the downfolding in Sec. IV. Summary is given in Sec.V.

\section{MODEL}\label{sec:model}
Here, we describe basic procedures for the derivation of effective low-energy models for the two compounds. The basis of the Hamiltonian is the Wannier function associated with anti-bonding/bonding states of HOMO or LUMO of dmit or BEDT-TTF molecules that form a dimer. The derived Hamiltonian is explicitly given in a form of the 2D multiband extended Hubbard model as 
\begin{eqnarray}
 \mathcal{H}&=& 
 \sum_{\sigma} \sum_{i} \sum_{\mu} \epsilon_{i\mu} 
 a_{i\mu\sigma}^{\dagger} a_{i\mu\sigma} +
 \sum_{\sigma} \sum_{i\neq j} \sum_{\mu} t_{i\mu j\mu} 
 a_{i\mu\sigma}^{\dagger} a_{j\mu\sigma} \nonumber \\
 &+& \frac{1}{2} \sum_{\sigma\rho} \sum_{ij} \sum_{\mu\nu} 
 V_{i\mu j\nu} 
 a_{i\mu\sigma}^{\dagger} a_{j\nu\rho}^{\dagger} 
 a_{j\nu\rho} a_{i\mu\sigma}  \nonumber \\ 
 &+& \frac{1}{2}\!\sum_{\sigma\rho}\!\sum_{ij}\!\sum_{\mu\nu} 
 \!J_{i\mu j\nu}\!\! \left(\!
 a_{i\mu\!\sigma}^{\dagger}\!a_{j\nu\!\rho}^{\dagger}\!
 a_{i\mu\!\rho}\!a_{j\nu\!\sigma}
 \!\!+\!\!
 a_{i\mu\!\sigma}^{\dagger}\!a_{i\mu\!\rho}^{\dagger}\!
 a_{j\nu\!\rho}\!a_{j\nu\!\sigma} \!\!\right)
\label{H_Hub3}
\end{eqnarray}
with $a_{i\mu\sigma}^{\dagger}$ ($a_{i\mu\sigma}$) being a creation (annihilation) operator of an electron with spin $\sigma$ in the $\mu$th Wannier orbital localized at the $i$th dimer site. The $\epsilon_{i\mu}$ and $t_{i\mu j\mu}$\ parameters are given by
\begin{eqnarray}
\epsilon_{i\mu}=\langle \phi_{i\mu}|\mathcal{H}_\mathrm{KS}|\phi_{i\mu}\rangle 
\label{eps_ij} 
\end{eqnarray}
and 
\begin{eqnarray}
t_{i\mu j\mu}=\langle\phi_{i\mu}|\mathcal{H}_\mathrm{KS}|\phi_{j\mu} \rangle, 
\label{t_ij} 
\end{eqnarray}
respectively, with $| \phi_{i\mu} \rangle$=$a_{i\mu}^{\dagger}|0\rangle$ and $\mathcal{H}_\mathrm{KS}$ being an effective one-body Kohn-Sham Hamiltonian. In this model for EtMe$_3$Sb[Pd(dmit)$_2$]$_2$, we take three Wannier orbitals, i.e., those of LUMO-b, HOMO-ab, and LUMO-ab, to represent the Hamiltonian.  In $\kappa$-(BEDT-TTF)$_2$Cu(NCS)$_2$, we take two Wannier orbitals; HOMO-ab and HOMO-b. Note that the Wannier orbitals are made for each band individually, since the bands separate with each other. With this construction of the Wannier function, the transfers for different orbitals are zero (i.e., $t_{i\mu j\nu}$=0 for $\mu$$\ne$$\nu$). The $V_{i\mu j\nu}$ and $J_{i\mu j\nu}$ parameters in Eq.~(\ref{H_Hub3}) are screened Coulomb and exchange integrals in the Wannier-orbital basis, respectively, expressed as 
\begin{eqnarray}
V_{i\mu j\nu}\!=\!\int\!\!\!\int d{\bf r} d{\bf r}'\!\! 
\phi_{i\mu}^{*}({\bf r}) \phi_{i\mu}({\bf r}) W({\bf r},{\bf r}') \phi_{j\nu}^{*}({\bf r}') \phi_{j\nu}({\bf r}') 
\label{V_ij}
\end{eqnarray} 
and
\begin{eqnarray}
J_{i\mu j\nu}\!=\!\int\!\!\!\int d{\bf r} d{\bf r}'\!\!
\phi_{i\mu}^{*}({\bf r}) \phi_{j\nu}({\bf r}) W({\bf r},{\bf r}') \phi_{j\nu}^{*}({\bf r}') \phi_{i\mu}({\bf r}') 
\label{J_ij}
\end{eqnarray}
with $W({\bf r},{\bf r}')$ being a 2D screened Coulomb interaction in the low-frequency limit.  Here, $V_{i\mu i\mu}$ is nothing but the onsite Hubbard $U$, while $J_{i\mu i\mu}$ is set to zero by definition. If one keeps only the HOMO-ab band crossing the fermi level and treats other bands as screening bands, the multi-band model is reduced to a single-band model 
\begin{eqnarray}
 \mathcal{H} 
&=& \sum_{\sigma} \sum_{i\neq j} t_{ij} a_{i \sigma}^{\dagger} a_{j \sigma}
 + \frac{1}{2} \sum_{\sigma \rho} \sum_{ij} V_{ij} a_{i \sigma}^{\dagger} 
   a_{j \rho}^{\dagger} a_{j \rho} a_{i \sigma}  \nonumber \\ 
&+& \frac{1}{2} \sum_{\sigma \rho} \sum_{i\neq j} J_{ij} \left(a_{i \sigma}^{\dagger} a_{j \rho}^{\dagger} a_{i \rho} a_{j \sigma} + a_{i \sigma}^{\dagger} a_{i \rho}^{\dagger} a_{j \rho} a_{j \sigma}\right), 
\label{H_Hub}
\end{eqnarray}
where we drop the orbital index $\mu$. 

The calculation for $W({\bf r},{\bf r}')$ in Eqs.~(\ref{V_ij}) and (\ref{J_ij}) follows Ref.~\onlinecite{Nakamura2D}, where a new cRPA framework was developed for the purpose to derive effective interactions of models defined in lower spatial dimensions. This new scheme is suitable for quasi-low-dimensional materials as the present system. The cRPA method is originally formulated in the RPA framework with the constraint for the band degree of freedom to eliminate only the degrees of freedom far from the fermi level in energy. This is called the band downfolding. In the proposed scheme of the supplementary downfolding, however, the concept of the constraint is additionally relaxed to include the screening by the polarization in the other layers/chains even within the target bands. This utilizes the real-space representation of the polarization function and elimination of specific-polarization blocks associated with the spatial dimensions perpendicular to the target layer/chain, which leads to a low-dimensional effective interaction for the target layer/chain. We call it the dimensional downfolding. Practically, the band+dimensional downfolding is performed in two steps: We first perform the band downfolding to derive a three-dimensional model for a small number of bands near the fermi level.~\cite{Nakamura3D} This is followed by the dimensional downfolding in the second step.~\cite{Nakamura2D} With this idea, we can naturally derive the low-energy model in any dimension. In the present case, we use it for the derivation of multiband and single-band 2D models for EtMe$_3$Sb[Pd(dmit)$_2$]$_2$ and $\kappa$-(BEDT-TTF)$_2$Cu(NCS)$_2$.

\section{Results}\label{sec:results}

\subsection{Computational Detail}
{\em Ab initio} density-functional calculations were performed by the program package xTAPP, which is a massively parallelized version of {\em Tokyo Ab initio Program Package} (TAPP) (Ref.~\onlinecite{TAPP}) and is able to perform the present large scale calculations efficiently. The xTAPP program adopts plane-wave basis sets and the present calculations were performed with norm-conserving pseudopotentials~\cite{PP} and generalized gradient approximation (GGA) (Ref.~\onlinecite{PBE96}) for the exchange-correlation potential. The experimental structure of EtMe$_3$Sb[Pd(dmit)$_2$]$_2$ at 4 K was taken from x-ray crystallography,~\cite{structure-dmit} while the atomic structure of $\kappa$-(BEDT-TTF)$_2$Cu(NCS)$_2$ at 15 K was taken from neutron data.~\cite{Schultz} For both of the systems, the positions of the hydrogen atoms were relaxed. The cutoff energies in wavefunctions and charge densities were set to 36 Ry and 256 Ry, respectively. 5$\times$5$\times$3 and 5$\times$5$\times$5 $k$-point samplings were employed for EtMe$_3$Sb[Pd(dmit)$_2$]$_2$ and $\kappa$-(BEDT-TTF)$_2$Cu(NCS)$_2$, respectively. The construction of MLWO follows Ref.~\onlinecite{MaxLoc}. The polarization function was expanded in plane waves with an energy cutoff of 5 Ry and the total number of bands considered in the polarization calculation was set to 2000 for EtMe$_3$Sb[Pd(dmit)$_2$]$_2$ and 750 for $\kappa$-(BEDT-TTF)$_2$Cu(NCS)$_2$. This condition corresponds to considering the excitation up to $\sim$20 eV above the fermi level. The Brillouin-zone (BZ) integral on the wavevector in the polarization calculation was evaluated by the generalized tetrahedron method.~\cite{Fujiwara} In the dimensional downfolding procedure, the calculated 2D effective Coulomb/exchange integrals were extrapolated to values with the infinite number of screening layers.~\cite{Nakamura2D,Shinaoka}

\subsection{EtMe$_3$Sb[Pd(dmit)$_2$]$_2$}
\begin{figure}[htbp]
\vspace{0cm}
\begin{center}
\includegraphics[width=0.5\textwidth]{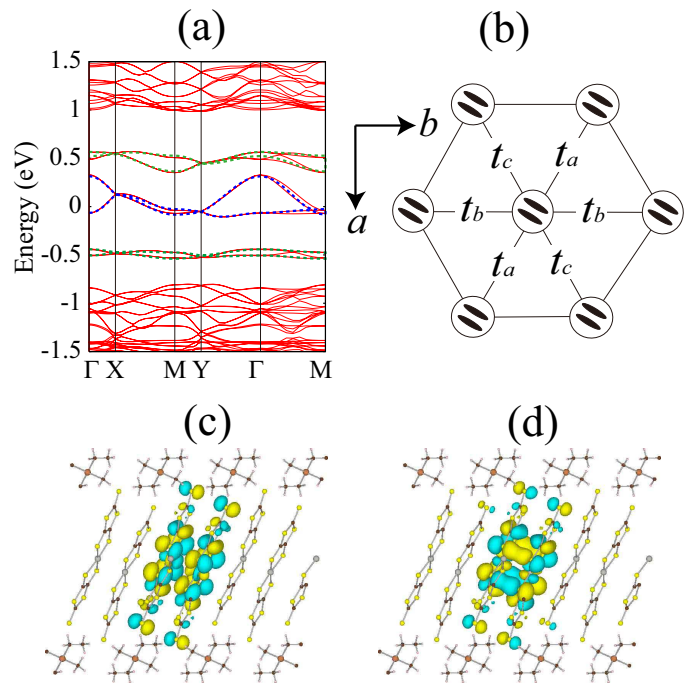}
\caption{(Color online) (a) Our calculated GGA band structure (solid lines) and the Wannier interpolated bands (dotted lines) with the transfers listed in Table~\ref{PARAM-dmit} of EtMe$_3$Sb[Pd(dmit)$_2$]$_2$. The crystal structure consists of alternating layers (parallel to the $ab$ plane) of the Pd(dmit)$_2$ anion and polymeric EtMe$_{3}$Sb$^{+}$ cation. Band dispersions are plotted along the high-symmetry points in the $ab$ plane, where $\Gamma$=(0, 0, 0), X=($a^{*}$/2, 0, 0), M=($a^{*}$/2, $b^{*}$/2, 0), Y=(0, $b^{*}$/2, 0). The zero of energy is the fermi level. The HOMO-ab band is drawn by the dark blue dotted lines and the LUMO-b and LUMO-ab bands are described by light green dotted lines. (b) Triangular lattice and the definition of transfers. The system contains two equivalent Pd(dmit)$_2$ layers and the dimer units stack along the $a$+$b$ direction in one layer and the $a$$-$$b$ direction in the other layer. In this figure, the latter-layer configuration is depicted. The $t_a$ direction is defined as the stacking direction and the $t_b$ direction is defined to be parallel to the $b$ axis. The $t_c$ direction is given as the direction of the remaining side of the triangular lattice. (c) Our calculated Wannier function of the HOMO-ab state in view along the $b$ axis, drawn by VESTA.~\cite{VESTA} The dark blue surfaces indicate positive isosurface and the light yellow surfaces indicate negative isosurface. (d) The same plot for the LUMO-b state.}
\label{Fig1}
\end{center}
\end{figure}

Figure~\ref{Fig1}~(a) shows our calculated GGA band structures (solid lines) of EtMe$_3$Sb[Pd(dmit)$_2$]$_2$. The band in [$-$0.1 eV: 0.3 eV] is the HOMO-ab band which is sandwiched by the LUMO-b and LUMO-ab bands at $\sim$$\pm$0.5 eV. The dotted lines represent the tight-binding band obtained with the transfer parameters in Table~\ref{PARAM-dmit}, where the definition of each transfer is given in the panel (b). The transfer parameters were estimated from the MLWOs. In the panels (c) and (d), we display our calculated MLWOs for the HOMO-ab and LUMO-b bands, respectively.  As the initial guess for the HOMO-ab Wannier orbital, we used the $p$-type Gaussian with the width 1.67 $\AA$, centered at the dimer center. Also, the initial guess for the LUMO-b and LUMO-ab MLWOs are superposition of the $p$-type Gaussians with width 1.18 $\AA$ put on the S atoms adjacent to Pd in each dmit molecule. 
\begin{table}[h] 
\caption{List of transfer parameters in the extended Hubbard model of EtMe$_3$Sb[Pd(dmit)$_2$]$_2$. For the definition of the transfers, see Fig.~\ref{Fig1}(b) and $\epsilon$ is the single-particle level energy. The unit is meV.} 

\centering 
\begin{tabular}{lr@{\ \ \ \ \ \ }r@{\ \ \ \ \ \ }r@{\ \ \ \ \ \ }r} 
\hline \hline
 & \multicolumn{1}{c}{$\epsilon$} & \multicolumn{1}{c}{$t_a$} &  \multicolumn{1}{c}{$t_b$} & \multicolumn{1}{c}{$t_c$}  \\ \hline \\ [-8pt] 
   LUMO-b   &$-$490 & 0       &  8.4  & $-$16.1 \\ 
   HOMO-ab  &34.8   & 54.4    & 44.9  &    40.2 \\  
   LUMO-ab  &495    & $-$25.7 & 24.8  &    15.1 \\ 
   HOMO-ab exH\"uckel & - & 28.2 & 27.0 & 25.0 \\ \hline \hline
\end{tabular} 
\label{PARAM-dmit} 
\end{table}

 The transfer parameters for the HOMO-ab band in the extended H\"uckel method are given in the bottom of the table.~\cite{structure-dmit} Note that, the notation for the transfers in Ref.~\onlinecite{structure-dmit} is different from the present paper; in the correspondence between the former and latter, $t_B$=$t_a$ $t_s$=$t_b$, and $t_r$=$t_c$. In comparison with the {\it ab initio} values, the H\"uckel values are smaller than the {\it ab initio} ones by the factor of 1/2-2/3. On top of that, the {\it ab initio} parameters exhibit an appreciable one-dimensional anisotropy along the $t_a$ direction, i.e., $t_a$$>$$t_b$$\sim$$t_c$. In contrast, the H\"uckel parameters is close to the equilateral triangular lattice as $t_a$$\sim$$t_b$$\sim$$t_c$. 

\begin{figure}[htbp]
\vspace{0cm}
\begin{center}
\includegraphics[width=0.5\textwidth]{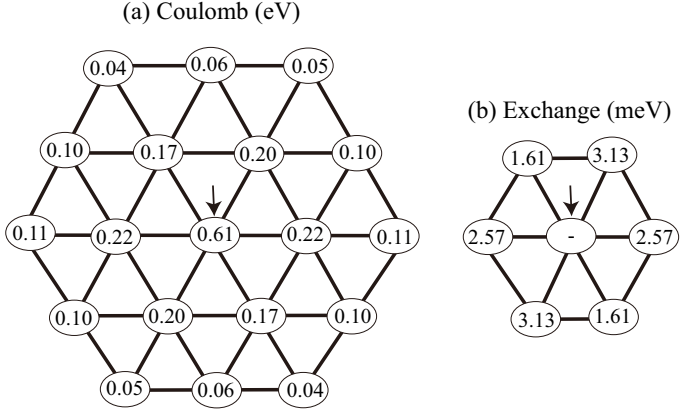}
\caption{Effective Coulomb (a) and exchange (b) integrals of the single band model for EtMe$_3$Sb[Pd(dmit)$_2$]$_2$ in Eq.~(\ref{H_Hub}). The arrow indicates a reference site. The view of the triangular lattice follows Fig.~\ref{Fig1}(b).} 
\label{Fig2}
\end{center}
\end{figure}

We next show in Fig.~\ref{Fig2} our calculated interaction parameters for the 2D single-band model. The panels (a) and (b) give the diagonal and off-diagonal parts of the interaction, respectively. In the presentation, we define a reference site as a site indicated by the arrow. The value in the ellipsoid represents the interaction value between the reference site and this ellipsoid one, in the unit of eV. For example, in the panel (a), 0.61 (eV) in the reference site is the onsite Hubbard $U$ and 0.22 (eV) next on the right is the offsite Coulomb integral between the reference site and this site. In the figure, we show the interaction value up to the tenth neighbors and the further distant interactions are negligibly small. Note that the system has a C$_2$ symmetry. The offsite exchange integral in the panel (b) is given in the unit of meV. In principle, the interaction range is short as up to the nearest shell and the values are so small as nearly half of those of $\kappa$-(BEDT-TTF)$_2$Cu(NCS)$_2$.~\cite{Shinaoka}  

\begin{figure}[htbp]
\vspace{0cm}
\begin{center}
\includegraphics[width=0.5\textwidth]{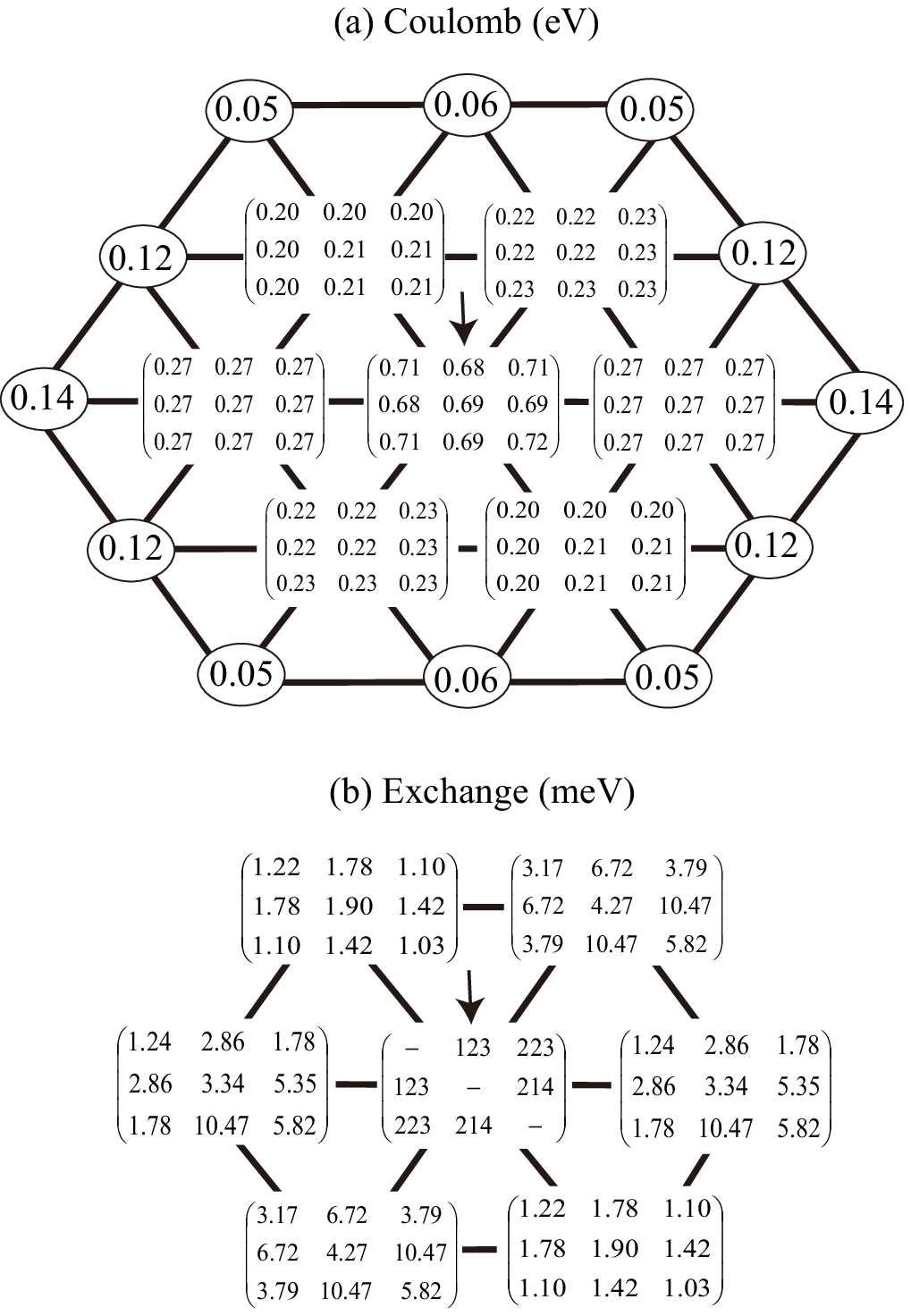}
\caption{Effective Coulomb (a) and exchange (b) integrals of the three-band model for EtMe$_3$Sb[Pd(dmit)$_2$]$_2$ in Eq.~(\ref{H_Hub3}). The arrow indicates a reference site. The view of the trianglar lattice follows Fig.~\ref{Fig1}(b).} 
\label{Fig3}
\end{center}
\end{figure}

The interaction values of the 2D three-band model are displayed in Fig.~\ref{Fig3}. In the multi-band case, the interaction is given as the 3$\times$3 matrix as 
\begin{eqnarray}
\label{eq-t1}
\left(
\begin{array}{ccc}
\phantom{ } & \phantom{VVVV} & \phantom{ } \\[-4mm]
V_{01j1} & V_{01j2} & V_{01j3}   \\
V_{02j1} & V_{02j2} & V_{02j3}   \\
V_{03j1} & V_{03j2} & V_{03j3}   \\
\end{array}
\right)
{\rm and} 
\left(
\begin{array}{ccc}
\phantom{ } & \phantom{VVVV} & \phantom{ } \\[-4mm]
J_{01j1} & J_{01j2} & J_{01j3}   \\
J_{02j1} & J_{02j2} & J_{02j3}   \\
J_{03j1} & J_{03j2} & J_{03j3}   \\
\end{array}
\right), 
\end{eqnarray}
where $V_{0\mu j\nu}$ ($J_{0\mu j\nu}$) indicates the Coulomb (exchange) integral between the $\mu$th orbital in the $0$th reference site and the $\nu$th orbital in the $j$th site. In the orbital labels, $(1,2,3) :=$ (LUMO-b, HOMO-ab, LUMO-ab). The onsite Coulomb interactions are nearly $\sim$0.7 eV and hold $U$$\sim$$U'$. On the offsite interactions, the orbital dependence is negligible, thus being well described as the function of the distance between the centers of the Wannier functions. On the basis of this aspect, for the interactions beyond the first shell, we show the averaged value over the orbitals.  The Hund's rule coupling at the reference site is rather large (e.g., the exchange integral between the LUMO-b and LUMO-ab orbitals as $\sim$0.22 eV). The exchange integrals with the neighboring site are rather orbital dependent, and there exist rather large values more than 10 meV. 

\subsection{$\kappa$-(BEDT-TTF)$_2$Cu(NCS)$_2$}
\begin{figure}[htbp]
\vspace{0cm}
\begin{center}
\includegraphics[width=0.45\textwidth]{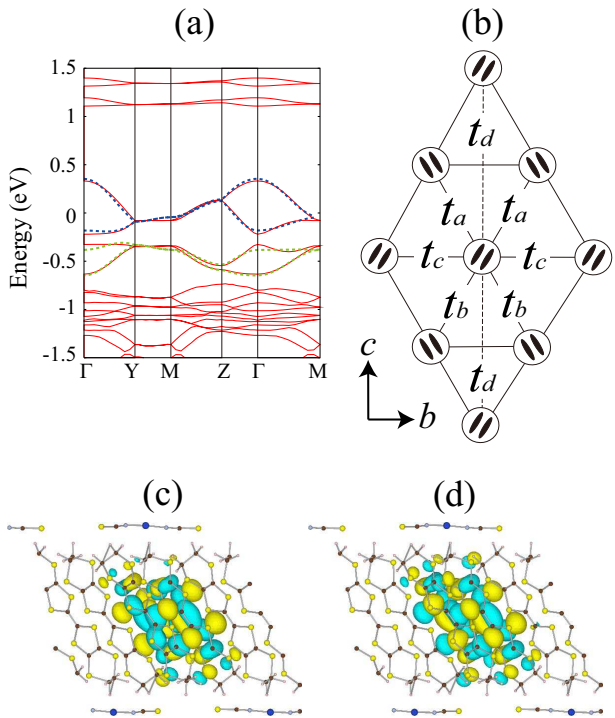}
\caption{(Color online) (a) Our calculated GGA band structure (solid lines) and the Wannier interpolated bands (dotted lines) with the transfers listed in Table~\ref{PARAM-BEDT} of $\kappa$-(BEDT-TTF)$_2$Cu(NCS)$_2$. The crystal structure contains alternating layers (parallel to the $bc$ plane) of BEDT-TTF donor molecules and polymeric Cu(NCS)$_{2}^{-}$ anions. Band dispersions are plotted along the high-symmetry points in the $bc$ plane, where $\Gamma$=(0, 0, 0), Y=(0, $b^{*}$/2, 0), Z=(0, 0, $c^{*}$/2), and M=(0, $b^{*}$/2, $c^{*}$/2). Note that the $a$ axis is interlayer axis. The zero of energy is the fermi level. The HOMO-ab band is drawn by the dark blue dotted lines and the HOMO-b band is described by light green dotted lines. (b) Definition of the transfers. The directions of the $b$ and $c$ axes forming the plane are also drawn. (c) Our calculated Wannier function of the HOMO-b state in view along the $a$ axis, drawn by VESTA.~\cite{VESTA} The dark blue surfaces indicate positive isosurface and the light yellow surfaces indicate negative isosurface. (d) The same plot for the HOMO-ab state.}
\label{Fig4-BEDT}
\end{center}
\end{figure}

We next move to $\kappa$-(BEDT-TTF)$_2$Cu(NCS)$_2$. Figure~\ref{Fig4-BEDT}~(a) displays our calculated GGA band (solid lines) and the tight-binding band (dotted lines) with the transfer integrals in Table~\ref{PARAM-BEDT}. The definition of each transfer follows the panel (b) and the Wannier function of the HOMO-b and HOMO-ab are plotted in the panels (c) and (d), respectively. The extended H\"uckel transfers for the HOMO-ab band~\cite{Kuroki,Komatsu} is given in the bottom of Table~\ref{PARAM-BEDT}. In $\kappa$-(BEDT-TTF)$_2$Cu(NCS)$_2$, $t_a$$\sim$$t_b$ holds and the lattice is approximated as an isosceles triangle with $t$=$(t_a+t_b)/2$ and $t'$=$t_c$. For both the {\it ab initio} and extended H\"uckel results, this trend well holds, but the former frustration strength $t'/t$ as $\sim$0.66 is somewhat smaller than the latter one as $\sim$0.84. We also note that, in the {\it ab initio} results, the $t_d$ transfer has a finite magnitude.
\begin{table}[h] 
\caption{List of transfer parameters in the extended Hubbard model of $\kappa$-(BEDT-TTF)$_2$Cu(NCS)$_2$. For the definition of the transfers, see Fig.~\ref{Fig4-BEDT}(b). The unit is meV.} 

\centering 
\begin{tabular}{lr@{\ \ \ }r@{\ \ \ }r@{\ \ \ }r@{\ \ \ }r} 
\hline \hline
 & \multicolumn{1}{c}{$\epsilon$} & \multicolumn{1}{c}{$t_a$} &  \multicolumn{1}{c}{$t_b$} & \multicolumn{1}{c}{$t_c$} & \multicolumn{1}{c}{$t_d$} \\ 
\hline \\ [-8pt] 
HOMO-ab  &0.02   & $-$64.8 & $-$69.3  &    44.2 & $-$11.5 \\  
HOMO-b   &$-$447 & $-$40.3 & $-$25.9  & $-$45.8 & 12.1    \\ 
HOMO-ab exH\"uckel & -\  & $-$69.3 & $-$65.8 & 56.6 &- \ \\ \hline \hline
\end{tabular} 
\label{PARAM-BEDT} 
\end{table}

\begin{figure}[htbp]
\vspace{0cm}
\begin{center}
\includegraphics[width=0.45\textwidth]{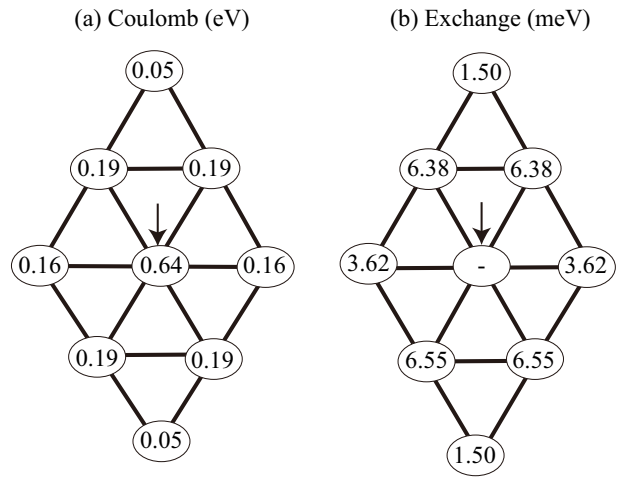}
\caption{Effective Coulomb (a) and exchange (b) integrals of the single band model for $\kappa$-(BEDT-TTF)$_2$Cu(NCS)$_2$ in Eq.~(\ref{H_Hub}). The arrow indicates a reference site. The view of the triangular lattice follows Fig.~\ref{Fig4-BEDT}(b).}
\label{Fig5-BEDT}
\end{center}
\end{figure}

The interaction parameter for the 2D single-band model is given in Fig.~\ref{Fig5-BEDT}, where the panels (a) and (b) are diagonal and off-diagonal parts for the interaction, respectively. How to see the figure is the same as Fig.~\ref{Fig2}. The interaction range for $\kappa$-(BEDT-TTF)$_2$Cu(NCS)$_2$ is shorter than that of EtMe$_3$Sb[Pd(dmit)$_2$]$_2$ as far as we compare the distance with the values up to $\sim$0.05 eV. This is due to the fact that the interdimer distance of $\kappa$-(BEDT-TTF)$_2$Cu(NCS)$_2$ is larger than that of EtMe$_3$Sb[Pd(dmit)$_2$]$_2$ [7.68 $\AA$ for the $\kappa$-BEDT-TTF salt and 6.31 $\AA$ for the Pd(dmit)$_2$ salt]. Furthermore, in general, the in-plane interaction range is roughly determined by the interlayer distance,~\cite{Nakamura2D,Shinaoka} because, in the dimensional downfolding, the interlayer metallic screening is switched on from this distance [16.44 $\AA$ for the $\kappa$-BEDT-TTF salt and 18.52 $\AA$ for the Pd(dmit)$_2$ salt] and, within this range, the in-plane interaction gives finite values. We also mention that the exchange value of $\kappa$-(BEDT-TTF)$_2$Cu(NCS)$_2$ is nearly twice as large as that of EtMe$_3$Sb[Pd(dmit)$_2$]$_2$, which may reflect the difference in the dimer orientation of $\kappa$-(BEDT-TTF)$_2$Cu(NCS)$_2$ and EtMe$_3$Sb[Pd(dmit)$_2$]$_2$ [Compare Fig.~\ref{Fig1}(b) and Fig.~\ref{Fig4-BEDT}(b)]. 

\begin{figure}[htbp]
\vspace{0cm}
\begin{center}
\includegraphics[width=0.5\textwidth]{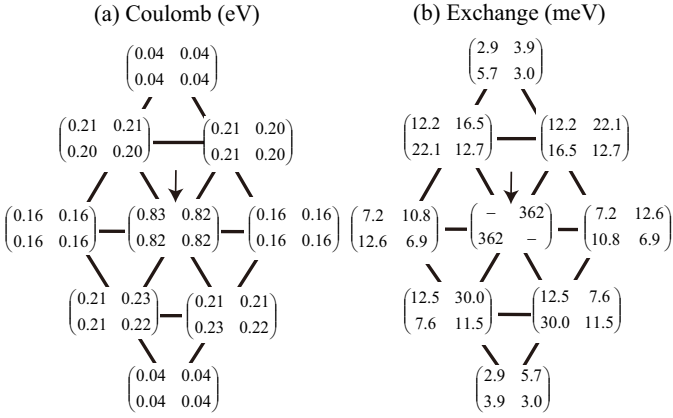}
\caption{Effective Coulomb (a) and exchange (b) integrals of the two band model for $\kappa$-(BEDT-TTF)$_2$Cu(NCS)$_2$ in Eq.~(\ref{H_Hub3}). The arrow indicates a reference site. The view of the triangular lattice follows Fig.~\ref{Fig4-BEDT}(b).} 
\label{Fig6-BEDT}
\end{center}
\end{figure}

We next show in Fig.~\ref{Fig6-BEDT} our calculated 2D interaction parameters for the two-band model. The interactions are given by the matrix form as  
\begin{eqnarray}
\label{eq-t1}
\left(
\begin{array}{cc}
\phantom{ } & \phantom{VV} \\[-4mm]
V_{01j1} & V_{01j2}  \\
V_{02j1} & V_{02j2}  \\
\end{array}
\right)
{\rm and} 
\left(
\begin{array}{ccc}
\phantom{ } & \phantom{VV} \\[-4mm]
J_{01j1} & J_{01j2}   \\
J_{02j1} & J_{02j2}   \\
\end{array}
\right).
\end{eqnarray}
In the orbital labels, $(1,2) :=$ (HOMO-ab, HOMO-b). The notations are the same as Fig.~\ref{Fig3}, but note that the symmetry of the interaction matrix is different from that for EtMe$_3$Sb[Pd(dmit)$_2$]$_2$. The onsite Coulomb interaction is $\sim$0.8 eV and the Hund's rule coupling is $\sim$0.36 eV, being larger than those of EtMe$_3$Sb[Pd(dmit)$_2$]$_2$. The offsite Coulomb integrals are not orbital dependent, while the offsite exchange integrals are largely orbital dependent and the largest value is as large as $\sim$30 meV. 

\section{Discussion}\label{sec:discussion} 
Here, we first remark that the antiferromagnetic exchange coupling $J$ is estimated in the strong-coupling perturbation as $J$ $\sim$ $4t^2$/$(U$$-$$V)$ = 245 K for the Pd(dmit)$_2$ salt and $J$ $\sim$ 360 K for $\kappa$-BEDT-TTF salt, when we use $t=(t_a+t_b+t_c)/3$ to compare with the experimental data deduced from the fitting to the isotropic triangular Heisenberg model in the high-temperature expansion.~\cite{tp-over-t-dmit-1}  When we consider the insufficiency of the single-band model and the experimental uncertainty coming from the lack of the data above 300 K, the results are favorably compared with 250 K estimated for the both compounds from such fittings of the magnetic susceptibility. 
 
\begin{figure}[htbp]
\vspace{0cm}
\begin{center}
\includegraphics[width=0.4\textwidth]{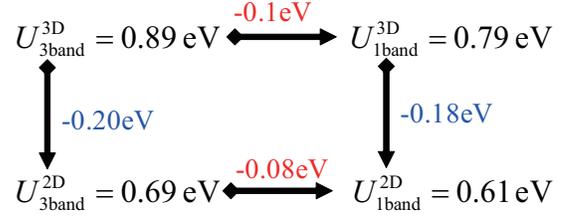}
\caption{(Color online) A schematic diagram of the screening process of the onsite interaction for the HOMO-ab Wannier orbital of EtMe$_3$Sb[Pd(dmit)$_2$]$_2$, generated by the band+dimensional downfolding.}
\label{Fig7}
\end{center}
\end{figure}

Now we discuss the screening effects due to the band downfolding from the multiband to single-band models and due to the dimensional downfolding as well. In Fig.~\ref{Fig7}, we draw a schematic diagram showing how the onsite interaction value of the HOMO-ab band of EtMe$_3$Sb[Pd(dmit)$_2$]$_2$ is screened. The bare Coulomb interaction is 3.55 eV and in the stage of the 3D three-band model with the conventional band downfolding, the interaction value becomes 0.89 eV. With considering the dimensional downfolding, the value is reduced further to 0.69 eV by 0.2 eV. When we consider the screening due to the LUMO-b and LUMO-ab bands in addition, to construct the single-band model, the value becomes 0.61 eV (the reduction of 0.08 eV). So, the largest part of the screening comes from the high-energy-electron screening considered in the initial band-downfolding stage, while the screening effects due to the dimensional downfolding, as well as the reduction to a single-band model may not be neglected quantitatively. We note that the fully-screened-RPA value is 0.14 eV.

\begin{figure}[htbp]
\vspace{0cm}
\begin{center}
\includegraphics[width=0.4\textwidth]{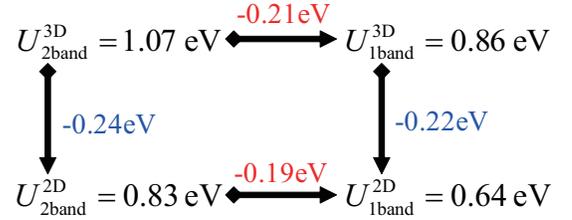}
\caption{(Color online) A schematic diagram of change in the onsite interaction for the HOMO-ab Wannier orbital of $\kappa$-(BEDT-TTF)$_2$Cu(NCS)$_2$, generated by the band+dimension downfolding.}
\label{Fig8}
\end{center}
\end{figure}

The same diagram for $\kappa$-(BEDT-TTF)$_2$Cu(NCS)$_2$ is shown in Fig.~\ref{Fig8}. In this material, while the reduction of the value due to the dimensional downfolding is nearly the same as the EtMe$_3$Sb[Pd(dmit)$_2$]$_2$ case, the reduction due to the band downfolding from the two-band to single-band model is appreciable (the reduction of $\sim$0.2 eV). This is probably due to the fact that the spatial distribution of the eliminated HOMO-b orbital is similar to that of the target HOMO-ab orbital and largely overlaps, resulting in a more efficient screening. The bare and fully-screened-RPA values are 3.61 and 0.19 eV, respectively. 

We next discuss the degree of the correlation strength of the 2D effective models, which is measured by ($U$$-$$V$)/$B$, where $U$ and $V$ are the onsite and offsite Coulomb interactions, respectively, and $B$ is the bandwidth. Since we are interested in low energy, we focus on the HOMO-ab band. Table~\ref{PARAM-WoverD} lists our calculated parameters and the correlation strength. For both EtMe$_3$Sb[Pd(dmit)$_2$]$_2$ and $\kappa$-(BEDT-TTF)$_2$Cu(NCS)$_2$, ($U$$-$$V$)/$B$ is comparable to the unity and thus these systems are classified as the strongly correlated electron system. Although the correlation strength of the single-band model is somewhat smaller than that of the multiband model, it is still large as close to 1. 
\begin{table*}[t] 
\caption{Our estimated correlation strength ($U$$-$$V$)/$B$ and downfolding measure $W'/D$. Here, $U$ and $V$ are onsite and offsite Coulomb interactions of the HOMO-ab Wannier orbitals, respectively, and $B$ is the bandwidth of the HOMO-ab band. Also, $W'$ is the matrix element of the full-RPA interaction between the HOMO-ab and LUMO-b Wannier orbitals for EtMe$_3$Sb[Pd(dmit)$_2$]$_2$ and HOMO-ab and HOMO-b Wannier orbitals for $\kappa$-(BEDT-TTF)$_2$Cu(NCS)$_2$ and $D$ is the minimum energy distance between the fermi level and bands away from the fermi level. The unit for $U$, $V$, $B$, $W'$, and $D$ are given in eV.} 

\ 

\centering 
\begin{tabular}{lr@{\ \ \ }r@{\ \ \ }r@{\ \ \ }r@{\ \ \ }r@{\ \ \ }r@{\ \ \ }r} 
\hline \hline
 & \multicolumn{1}{c}{$U$} & \multicolumn{1}{c}{$V$} & \multicolumn{1}{c}{$B$}  & \multicolumn{1}{c}{$\frac{U-V}{B}$} & \multicolumn{1}{c}{$W'$} 
 & \multicolumn{1}{c}{$D$} & \multicolumn{1}{c}{$\frac{W'}{D}$} \\ 
\hline \\ [-8pt] 
 EtMe$_3$Sb[Pd(dmit)$_2$]$_2$ (single-band) & 0.61 & 0.20 & 0.44 & 0.93 & -    & -    &  -   \\ 
 EtMe$_3$Sb[Pd(dmit)$_2$]$_2$ (three-band) & 0.69 & 0.23 & 0.44 & 1.04 & 0.15 & 0.44 & 0.33 \\ 
 $\kappa$-(BEDT-TTF)$_2$Cu(NCS)$_2$ (single-band) & 0.64 & 0.18 & 0.56 & 0.82 & -    & -    &  -   \\ 
 $\kappa$-(BEDT-TTF)$_2$Cu(NCS)$_2$ (two-band) & 0.83 & 0.19 & 0.56 & 1.14 & 0.16 & 0.33 & 0.49 \\ 

\hline \hline
\end{tabular} 
\label{PARAM-WoverD} 
\end{table*}

\begin{figure}[h]
\vspace{0cm}
\begin{center}
\includegraphics[width=0.5\textwidth]{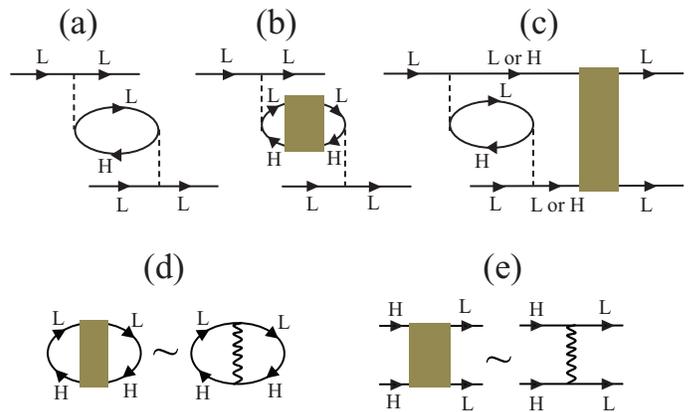}
\caption{(Color online) (a) A schematic diagram for effective interactions of the single-band model, where we draw symbolically the second-order one. In cRPA, the interaction between low-energy electrons in the HOMO-ab band (external lines) is screened by polarizations (internal lines) via high-energy electrons, where the capital characters ``L" and ``H" describe propagators of the low- and high-energy electrons, respectively. In the panel (b), we show the correction to the effective interaction (a), where the polarization is modified by the vertex (shaded area) ignoring the RPA process, while in (c), the vertex introduces an additional interaction process. In the first-order approximation, the vertex is replaced by the fully screened interaction. We draw the schematic diagrams for those; (d) the vertex correction to the polarization and (e) the ladder-type interaction for the propagators.}
\label{Fig9}
\end{center}
\end{figure}

We now detail the validity of reducing from the multiband to single-band model in the present case, by applying a general criterion.~\cite{ImadaMiyake} In cRPA for deriving the effective interaction of the single-band model, the multiband is divided into the target HOMO-ab band and the rest. Then, with excluding a polarization formed in the target band, the effective interaction is calculated. Figure~\ref{Fig9} (a) exhibits a schematic diagram for deriving the single-band effective interaction illustrated symbolically in the second order one, where low-energy electrons in the HOMO-ab band (external lines) is screened by a polarization (internal lines) via high-energy electrons (namely, electrons belonging to the bands away from the fermi level) in the eliminated bands [i.e., the LUMO-ab/LUMO-b bands for the Pd(dmit)$_2$ salt and the HOMO-b band for the $\kappa$-BEDT-TTF salt]. Here, the capital characters ``L" and ``H" describe propagators of the low- and high-energy electrons, respectively. In the figure, an electron-hole pair formed in the low- and high-energy bands is displayed. The infinite series of such a diagram gives the effective interaction used in the single-band model. In this treatment, the vertex which modifies the interaction is totally dropped. In the panels (b) and (c), we show the correction to the original interaction process in (a); the vertex in the shaded area corrects the polarization [panel (b)] or introduces an additional interaction process [(c)]. To the first approximation, the vertex is approximated by the single fully screened RPA interaction; the correction to the polarization is written as an exchange process (d) and the additional interaction process for the propagators is written as a single ladder process (e). 

Let us consider the magnitude of these corrections to the original process. The polarization is given by $-iG_{H}G_{L}$ with $G_{L}$ and $G_{H}$ being the propagators of the low- and high-energy electrons, respectively, and this magnitude roughly scales as $-(1/D)$, where $D$ is the minimum energy distance of the high-energy band from the fermi level. We note that, in this estimate, the effect of the transition matrix element is dropped for the simplicity. Similarly, the correction term described in the panel (d) is given by $G_{H}$$G_{L}$$G_{H}$$G_{L}$$W_{LH}\sim W_{LH}/D^2$ with $W_{LH}$ being the fully-screened interaction between the low- and high-energy electrons.~\cite{Note_WLH} Thus, the order of the correction to the original polarization is scaled as $W_{LH}$/$D$.~\cite{ImadaMiyake} From the similar discussion, the order of the correction described in the panel (e) is also given as $W_{LH}$/$D$. This value of $W_{LH}$/$D$ is a measure for a stable downfolding from a multiband to a single-band model. When this value is small, the vertex correction to the cRPA interaction can be neglected. On the contrary, when the value is large, the cRPA downfolding breaks down and one must solve the multiband model directly without reducing to the single-band model. 

We show in Table~\ref{PARAM-WoverD} our estimated $W_{LH}/D$, where we have approximated $W_{LH}$ by the full RPA form $W'$; in EtMe$_3$Sb[Pd(dmit)$_2$]$_2$, $W'$ is calculated as the full-RPA Coulomb interaction matrix element between the HOMO-ab and LUMO-b Wannier orbitals, and, in $\kappa$-(BEDT-TTF)$_2$Cu(NCS)$_2$, $W'$ is estimated as the matrix element between the HOMO-ab and HOMO-b Wannier orbitals. The resulting $W'/D$ is as large as 0.3-0.5. This value is rather large compared to the transition-metal oxides. In the transition-metal oxide, $d$ bands ($t_{2g}$ or $e_g$ bands) form the low-energy band and the closest high-energy bands consist of oxygen-$p$ bands. In this case, the matrix element of the full-RPA interaction $W_{dp}$ between $t_{2g}$/$e_g$ and O$_p$ orbitals is substantially smaller, because the centers of the Wannier orbitals are spatially apart with each other and the full-RPA interaction does not work for such a length scale. On the other hand, in the case of the organic compounds, the Wannier orbitals of the low- and high-energy bands have the same centers and therefore the matrix element for the full-RPA interaction does not disappear. This is a characteristic feature of the organic compound compared to the transition-metal oxide. As a result, in the case of the organic compound, the downfolding treatment based on cRPA to reduce to the single-band model would be insufficient and the multiband analysis is strongly recommended. 

One of the compounds exhibiting quantum spin-liquid behavior is $\kappa$-(BEDT-TTF)$_2$Cu$_2$(CN)$_3$. The {\it ab initio} parameters for the 3D single-band systems have been derived in Ref.\onlinecite{NakamuraET}.  Since the HOMO bonding orbital appears to be entangled with other lower-energy bands, one may need to treat more than two bands for the multi-band description, in contrast to $\kappa$-(BEDT-TTF)$_2$Cu(NCS)$_2$, which is left for future studies. 

\section{Conclusion}
To summarize, in the present paper, we derived the 2D multiband models for EtMe$_3$Sb[Pd(dmit)$_2$]$_2$ and $\kappa$-(BEDT-TTF)$_2$Cu(NCS)$_2$, with the {\it ab initio} cRPA plus MLWO framework suitable for strongly correlated electron systems. The correlation strength for the derived model, ($U$$-$$V$)/$B$, is estimated as a large value of $\sim$1, as shown in Table~\ref{PARAM-WoverD}, indicating the strongly correlated nature of electrons in these systems.  The transfer structure of EtMe$_3$Sb[Pd(dmit)$_2$]$_2$ is characterized by a lattice of quasi-2D scalene triangular lattice with 1D anisotropy, whereas that of $\kappa$-(BEDT-TTF)$_2$Cu(NCS)$_2$ approximately has a 2D square-lattice structure with a weaker next-nearest neighbor transfer. To discuss whether the multiband model can be reduced to the single-band model with the usual cRPA treatment, we estimated the order of the vertex correction to the cRPA interaction. We found that the vertex process between the target HOMO-ab band and others [i.e., the LUMO-ab/LUMO-b bands for the Pd(dmit)$_2$ salt and the HOMO-b band for the $\kappa$-BEDT-TTF salt] is not small. Therefore, the multiband model should directly be analyzed, not by reducing to the single-band model, to clarify the realistic correspondence with strongly correlated phenomena including non-fermi liquid behavior, quantum spin liquid phase and/or unconventional superconducting mechanism found in the real organic compounds. This remains to be explored. 

\begin{acknowledgements} 
We would like to thank Reizo Kato and Takeo Fukunaga for providing us with the structural data for EtMe$_3$Sb[Pd(dmit)$_2$]$_2$. We also thank Shiro Sakai for fruitful discussions. Calculations were done at Supercomputer center at Institute for Solid State Physics, University of Tokyo. This work was supported by Grants-in-Aid for Scientific Research (No.~22740215, 22104010, 23110708, 23340095, 22104006, 19051016) from MEXT, Japan. A part of this research has been funded by the Strategic Programs for Innovative Research (SPIRE), MEXT, and the Computational Materials Science Initiative (CMSI), Japan.
\end{acknowledgements}


\begin{thebibliography}{99}
\bibitem{Note_WLH} 
The magnitude of the fully-screened interaction $W_{LH}$ may roughly be approximated by inverse of the polarization of the low-energy band, ($-iG_LG_L$)$^{-1}$.
\end{thebibliography}
\end{document}